\documentclass[12pt]{article}
\usepackage{amsfonts,amsmath,amssymb}
\usepackage[pdftex]{graphicx}
\usepackage{hyperref}
\usepackage{float}
\usepackage[allcommands, old-arrows]{overarrows}

\newcommand{\pa}{\partial}
\newcommand{\nn}{\nonumber}
\newcommand{\Ord}{{\cal O}}

\makeatletter

\@addtoreset{equation}{section}
\makeatother

\def\href#1#2{#2}
\begin{document}

\begin{titlepage}

\begin{center}

\hfill 
\vskip18mm

\textsl{\LARGE The Internal Structure of the}\\[4mm] 
\textsl{\LARGE Deconstructed Dirac Monopole}\\[16mm] 

{\large Kazuyuki Furuuchi}\\[8mm]

{\it
{Manipal Centre for Natural Sciences}\\
{Manipal Academy of Higher Education}\\
{Manipal 576 104, Karnataka, India}
}

\vskip8mm 
\end{center}
\begin{abstract}
I study the internal structure of 
the Dirac magnetic monopole 
in a deconstructed $U(1)$ gauge theory.
The deconstructed $U(1)$
gauge theory
has a product $U(1)^N$ gauge group,
which breaks down to the diagonal $U(1)_{\mathrm{diag}}$ gauge group.
A linear superposition of the Dirac monopoles 
each from each $U(1)$ gauge group
placed on top of each other in the 3D space
constitutes a Dirac monopole 
with a unit magnetic charge under the unbroken $U(1)_{\mathrm{diag}}$ gauge group.
However, the Dirac monopole in each $U(1)$ gauge group 
has a fractional
magnetic charge of the unbroken $U(1)_{\mathrm{diag}}$ gauge group with the same sign,
which makes these ``constituent'' Dirac monopoles repel each other.
Therefore, for the ``composite'' Dirac monopole of the $U(1)_{\mathrm{diag}}$ gauge group
to be stable, there must be attractive forces
that counter the repulsive magnetic Coulomb forces.
I argue that such attractive forces are provided by the 
Nielsen-Olesen type magnetic flux tubes of unbroken gauge groups.
This internal structure of the
composite Dirac monopole of $U(1)_{\mathrm{diag}}$ gauge group 
resembles the composite magnetic monopole 
found in the model constructed by Saraswat \cite{Saraswat:2016eaz}.
I estimate the size of the Dirac monopole in $U(1)_{\mathrm{diag}}$ gauge group
from the balance between the magnetic Coulomb forces and 
the forces from the tension of the magnetic flux tubes.
Implications of the results for the Weak Gravity Conjecture are briefly discussed.
\end{abstract}

\end{titlepage}

\section{Introduction}\label{sec:intro}

Dimensional deconstruction 
\cite{Hill:2000mu,Arkani-Hamed:2001kyx}
provides a 4D QFT description of 
latticized extra dimensions.
Dimensional deconstruction
has been applied to solve the hierarchy problem in
the electro-weak gauge symmetry breaking
\cite{ArkaniHamed:2001nc}.
While a difficulty in applying
dimensional deconstruction to large-field inflation has been identified
\cite{ArkaniHamed:2003wu},
models to circumvent or overcome this difficulty have been proposed
\cite{Furuuchi:2020klq,Furuuchi:2020ery}.

Solitons and instantons play important roles
in non-perturbative aspects of gauge theories.
Hence it is of interest to analyze
solitons and instantons
in deconstructed gauge theories.
The Dirac magnetic monopole in deconstructed
$U(1)$ gauge theories has been studied
in \cite{Poppitz:2003uz}.
Deconstructed $U(1)$
gauge theores
have a product $U(1)^N$ gauge group,
which breaks down to diagonal $U(1)_{\mathrm{diag}}$ gauge group.
The solution discussed in \cite{Poppitz:2003uz}
is a simple linear superposition of the Dirac monopoles 
each from each $U(1)$ gauge group,
placed on top of each other in the 3D space.
However, the Dirac monopole in each $U(1)$ gauge group
has a fractional magnetic charge of the $U(1)_{\mathrm{ diag}}$ gauge group with the same sign,
therefore, there are repulsive magnetic Coulomb forces 
between those ``constituent'' Dirac monopoles.
Thus, to make the ``composite'' Dirac monopole of the
$U(1)_{\mathrm{diag}}$ gauge group stable,
there must be attractive forces to counter the
repulsive magnetic Coulomb forces.
One may naively think that
when the 
constituent Dirac monopoles are displaced from
each other,
the ``Kaluza-Klein (KK) modes'' 
in the latticized extra dimension
are excited,
providing the required attractive force.
Therefore, the Dirac monopole in $U(1)_{\mathrm{ diag}}$ gauge group
would be stabilized at the KK scale.
However, while it is correct that 
the gauge fields of broken gauge groups play roles
in stabilizing the Dirac monopole in $U(1)_{\mathrm{ diag}}$ gauge group,
the KK scale is not the scale
at which the Dirac monopole is stabilized.
This can be understood from the following simple argument:
One can consider the formal continuum limit $N \rightarrow \infty$
with a fixed KK scale.
In this limit, the deconstructed $U(1)$ gauge theory becomes
a $U(1)$ gauge theory in 5D,
and the Dirac monopole solution
becomes a ``monopole-loop'' solution,
which is a Dirac monopole in 4D space-time
that is invariant under translations in the compactified 5th direction.
The 5D $U(1)$ gauge theory can describe
the solution above the KK scale,
and nothing special happens to the monopole-loop solution at the KK scale.
Therefore, the KK scale is not the scale
at which the Dirac monopole in $U(1)_{\mathrm{ diag}}$ is stabilized.
The above argument also suggests that the lattice spacing
may be the scale at which the Dirac monopole in $U(1)_{\mathrm{ diag}}$ is stabilized.

In this article,
I will study the internal structure of the
Dirac monopole in the $U(1)_{\mathrm{ diag}}$ gauge group in some detail.
I identify the Nielsen-Olesen type magnetic flux tubes \cite{Nielsen:1973cs}
between the constituent Dirac monopoles \cite{Nambu:1974zg}
as the source of the attractive forces.
A similar configuration was
studied earlier by Saraswat \cite{Saraswat:2016eaz} (see also \cite{Furuuchi:2017upe})
in a model constructed to examine the validity of the Weak Gravity Conjecture (WGC) 
\cite{ArkaniHamed:2006dz} in effective field theories (EFTs).
In particular, 
the composite magnetic monopole (CMM) constructed in \cite{Saraswat:2016eaz}
plays a crucial role in validating the magnetic version of the WGC
\cite{Furuuchi:2017upe} in that model.
Indeed, the current work was largely inspired by
that work \cite{Saraswat:2016eaz} and 
my experience in the subsequent works \cite{Furuuchi:2017upe,Pathak:2025ukb}.
In particular, the $N=2$ case 
reduces to a slightly extended version of the model of Saraswat
with $Z=1$, where $Z$ is the charge of the Higgs field 
in one of two $U(1$ gauge groups in that model.
In the meantime, the dimensional deconstruction at large $N$ 
is quite different from the model of Saraswat.
By analyzing the balance between the 
repulsive magnetic Coulomb forces and
the attractive forces by the tension of the magnetic flux tubes, 
I show that the lattice spacing of the deconstructed dimension
is the length scale also in the 3D space
below which the internal structure of the
Dirac monopole of the $U(1)_{\mathrm{diag}}$ gauge group manifests itself.
This is quite expected,
as at the distance scale larger than the lattice spacing 
(but below the KK scale),
the approximate 5D Lorentz symmetry recovers.

While the main theme of the current work is 
somewhat different from
the works on the WGC
\cite{Saraswat:2016eaz,Furuuchi:2017upe},
I also briefly discuss
the implications of the internal structure of 
the deconstructed Dirac monopole in the WGC.

\section{Dimensional Deconstruction of 
a $U(1)$ Gauge Theory%
}\label{sec:model}

The model I study is an
Abelian gauge theory in 4D
plus the deconstructed dimension.
I remind the readers that this model is a 4D quantum field theory (QFT),
and the additional latticized dimension is in the field space of the 4D QFT.
However, in the limit where the number of the lattice points $N$ is taken large,
the model approaches to a 5D Abelian gauge theory compactified on a circle.

I start with the action \cite{Hill:2000mu}:
\begin{align}
S
=
\int 
& d^4x
\sum_{j=0}^{N-1}
\Biggl[
-\frac{1}{4}
F_{\mu\nu\,(j)}
F^{\mu\nu}_{(j)}
+
\frac{1}{2}
D_\mu H_{(j,j+1)}^\ast
D^\mu H_{(j,j+1)}
- V(H)+\cdots
\Biggr]\,.
\label{eq:S}
\end{align}
Here, $j$'s are integers mod $N$.
The action \eqref{eq:S} has a 
$\prod_{j=0}^{N-1} U(1)_{(j)}$ gauge symmetry.
The gauge transformation generated by
$g_{(j)}(x) = e^{i g \alpha_{(j)}(x)}$ is given as
\begin{align}
A_{\mu (j)} (x) &\rightarrow A_{\mu (j)}(x)- \pa_\mu \alpha_{(j)}(x)\,,\\
H_{(j,j+1)} (x) &\rightarrow g_{(j)}^{-1}(x) H_{(j,j+1)}(x) g_{(j+1)}(x)\,.
\label{eq:gaugetr}
\end{align}
I also impose an exact global symmetry under
the discrete translation:
\begin{equation}
j \rightarrow j+1 \,, \qquad (j=0,\cdots,N-1\mbox{ mod }N)\,.
\label{eq:dts}
\end{equation}

The dots in the action in \eqref{eq:S}
indicate renormalisable terms including charged matter fields,
subject to the discrete translational symmetry \eqref{eq:dts}.
However, in this article,
I am only interested in the smallest electric charge unit
which determines
the possible smallest magnetic charge via the Dirac quantization condition.
Therefore, I just declare that the gauge coupling $g$
is normalized so that the smallest electric charge unit in the theory is one.
Due to the discrete translational symmetry \eqref{eq:dts},
the gauge couplings for all the $U(1)_{j}$ groups 
($j=0,\cdots,N-1$) are equal.

The covariant derivatives
in \eqref{eq:S}
are defined as
\begin{align}
D_\mu H_{(j,j+1)}
&=
\pa_\mu H_{(j,j+1)} 
- i g A_{\mu (j)} H_{(j,j+1)} + i g H_{(j,j+1)} A_{\mu (j+1)} \,,
\label{eq:DH}
\end{align}

The potential term $V(H)$ is such that
the vacuum expectation values of $H_{(j,j+1)}$ is given as
\begin{equation}
\left| \langle H_{(j,j+1)} \rangle \right| = f\,, \qquad \mbox{for all $j$}\,. 
\label{eq:vac}
\end{equation}
To be concrete, I consider
\begin{equation}
V(H) = 
\sum_{j=0}^{N-1} 
\left[
- \frac{\mu^2}{2} \left| H_{(j,j+1)}\right|^2 + \frac{\lambda}{4} \left| H_{(j,j+1)}\right|^2 
\right]\,,
\label{eq:VH}
\end{equation}
with 
$\mu= \sqrt{\lambda}\, f$.
Here, only the renormalisable terms that satisfy the exact symmetries,
i.e. 4D Lorentz symmetry, 
the gauge symmetry \eqref{eq:gaugetr}, 
and the discrete translation symmetry \eqref{eq:dts}, are considered.
I decompose the Higgs fields around the vacuum \eqref{eq:vac}:
\begin{equation}
H_{(j,j+1)}(x)
= 
\left( f+h_{(j,j+1)}(x) \right) U_{j,j+1}(x) 
\,,
\qquad (j = 0, \cdots, N-1 \mbox{mod} N)
\,.
\end{equation}
The radial component $h_{(j,j+1)}$
has mass
\begin{equation}
m_h^2 = 2{\lambda f^2} = 2\mu^2\,,
\label{eq:mh}
\end{equation}
in the vacuum.

At the energy scales below $m_h$,
I can integrate out $h_{(j,jj+1)}$
and the action reduces to
\begin{align}
S_{\mathrm{dec}}
=
\int 
& d^4x
\sum_{j=0}^{N-1}
\Biggl[
-\frac{1}{4}
F_{\mu\nu\,(j)}
F^{\mu\nu}_{(j)}
+
\frac{f^2}{2}
D_\mu U_{(j,j+1)}^\ast
D^\mu U_{(j,j+1)}
+ \cdots
\Biggr]\,.
\label{eq:Sdec}
\end{align}
This is the action for the
dimensional deconstruction.
The action \eqref{eq:S} is regarded as a UV model
for the dimensional deconstruction of a $U(1)$ gauge theory.
The action \eqref{eq:Sdec} is non-renormalizable
and should be regarded as an effective field theory (EFT).
The perturbative expansions in this EFT break down at 
the energy scale 
\begin{equation}
\Lambda_{\mathrm{pert}} := 4 \pi f\,.
\label{eq:LambdaUV}
\end{equation}
Note that $\Lambda_{\mathrm{pert}} \gtrsim m_H$ in
the perturbative regime
$\lambda \lesssim \Ord(1)$. 
It means that the dimensional deconstruction EFT
is replaced by the new model below $\Lambda_{\mathrm{pert}}$
if the UV model is described by the action \eqref{eq:S}.
The UV model \eqref{eq:S} is renormalizable.
Hence, formally, its UV cut-off scale $\Lambda_{\mathrm{ UV}}$ can be taken to infinity.
However, the magnetic WGC predicts that if one couples this theory with gravity,
the following bound is imposed
\cite{ArkaniHamed:2006dz}:\footnote{%
I used the WGC bound for $N$ $U(1)$ gauge groups \cite{Cheung:2014vva}.}
\begin{equation}
\Lambda_{\mathrm{ UV}} \lesssim 
\frac{g M_P}{\sqrt{N}}\,.
\end{equation}
In this article, I assume
that this bound from the magnetic WGC is satisfied
(this will be relevant only in Sec.~\ref{sec:WGC}).

The gauge transformation 
of $U_{(j,j+1)}$ follows from \eqref{eq:gaugetr}:
\begin{equation}
U_{(j,j+1)} (x) \rightarrow g_{(j)}^{-1}(x) U_{(j,j+1)}(x) g_{(j+1)}(x)\,.
\label{eq:gaugetr2}
\end{equation}
The covariant derivatives for $U_{(j,j+1)}$ follows from \eqref{eq:DH}:
\begin{align}
D_\mu U_{(j,j+1)}
&=
\pa_\mu U_{(j,j+1)} 
- i g A_{\mu (j)} U_{(j,j+1)} + i g U_{(j,j+1)} A_{\mu (j+1)} \,.
\label{eq:DU}
\end{align}

While the action \eqref{eq:Sdec} defines a 4D QFT,
it has the same structure as
the lattice gauge theories
on a periodically identified lattice,
which effectively gives rise to a latticized extra dimension.
The lattice points are parametrized by the label 
$j=0,1,\cdots,N-1 \mbox{ mod } N$).
The field $U_{(j,j+1)}(x)$ takes values in $U(1)$
and can be parametrized as
\begin{equation}
U_{(j,j+1)}(x) 
= \exp \left[ i \frac{A_{(j,j+1)} (x)}{f} \right]
= \exp \left[ i g d A_{(j,j+1)} (x) \right]
\,,
\label{eq:Link}
\end{equation}
where the field $A_{(j,j+1)}$ is real and periodically identified
as $A_{(j,j+1)} \sim A_{(j,j+1)} + 2\pi f$.
Here, 
following the terminology in the lattice gauge theories,
I introduced the lattice spacing $d$ defined by
\begin{equation}
d := \frac{1}{gf} \,.
\label{eq:d}
\end{equation}
Since the lattice is periodically identified,
it is convenient 
to introduce the radius $L_{\mathrm{ KK}}$ of the 
latticized extra dimension as
\begin{equation}
2 \pi L_{\mathrm{ KK}} := N a\,.
\label{eq:LKK}
\end{equation}

Next, I consider the action of the
EFT which is appropriate
below the KK energy scale $1/L_{\mathrm{ KK}}$:
\begin{align}
S_4
=
\int d^4x
\Biggl[
&-
\frac{1}{4}
F_{\mu\nu}
F^{\mu\nu}
+
\frac{1}{2}
\pa_\mu A
\pa^\mu A
+ \cdots
\Biggr]\,.
\label{eq:S4}
\end{align}
The fields appearing in the action \eqref{eq:S4}
are the ``zero-modes'' in the KK modes
from the latticized extra dimension:
\begin{align}
A_\mu 
&= 
\tilde{A}_{\mu\, (0)}
=
\frac{1}{\sqrt{N}}
\sum_{j=0}^{N-1}
A_{\mu\,(j)}\,,
\label{eq:Amuzero}\\
A 
&= 
\tilde{A}_{(0)}
=
\frac{1}{\sqrt{N}}
\sum_{j=0}^{N-1}
A_{(j,j+1)}\,.
\label{eq:Azero}
\end{align}
Appendix~\ref{sec:aDFT} summarizes the notation 
for the Discrete Fourier Transform (DFT)
used in this article.

I will call the EFT described by the action \eqref{eq:S4}
as ``Low Energy EFT,''
which is the low-energy effective theory of the
``High Energy EFT'' described by the action \eqref{eq:S}.
As mentioned earlier, the gauge coupling $g$ of the High Energy EFT
is normalized so that the smallest charge unit in the theory is one.
Below the KK scale, 
I also normalize the gauge coupling $g_4$
so that the smallest charge unit
in the Low Energy EFT is one:
\begin{equation}
g_4 = \frac{g}{\sqrt{N}} \,.
\label{eq:g4}
\end{equation}
To understand this normalization,
it is convenient to introduce ortho-normal basis vectors
for the original $U(1)_{(j)}$ gauge groups
and the ortho-normal basis vectors
that correspond to the mass-eigen states, as described below.

Of the product gauge group 
$\prod_{j=0}^{N-1} U(1)_{(j)}$
of the High Energy EFT,
only the diagonal part $U(1)_{\mathrm{diag}}$
gauge group is unbroken.
It is convenient to
use the basis in the gauge fields
in which the diagonal $U(1)_{\mathrm{diag}}$ and 
other broken $U(1)$ gauge groups are orthogonal.
The convenient basis is provided by the mass eigen-states, 
or Discrete Fourier Transform (DFT).
See Appendix~\ref{sec:aDFT} for the convention of DFT
in this article.

I first introduce the ortho-normal basis
in $N$-dimensional vector space
labeling the $U(1)_{(j)}$ gauge groups:
\begin{equation}
\vec{e}_{j}\,,
\quad \vec{e}_{j} \cdot \vec{e}_{\ell} = \delta_{j\ell}\,,
\quad
(j,\ell = 0,1,\cdots,N-1)\,.
\label{eq:ej}
\end{equation}
The metric for the inner product is 
determined by the kinetic term of the gauge fields
in \eqref{eq:Sdec},
which I assumed to be canonically normalized, 
for simplicity (no kinetic mixing).

The ortho-normal mass eigen-states basis is then given as
(the ortho-normality directly follows from \eqref{eq:DFTo})
\begin{align}
\tilde{\vec{e}}_{0}
&=
\frac{1}{\sqrt{N}}
\sum_{j=0}^{N-1}
\vec{e}_{j}\,,
\label{eq:ezero}\\
\tilde{\vec{e}}_{\frac{N}{2}}
&=
\frac{1}{\sqrt{N}}
\sum_{j=0}^{N-1}
(-)^j \vec{e}_{j}\,,
\quad \mbox{(when $N$ is even)}\,,\\
\tilde{\vec{c}}_{n} 
&= 
\sqrt{\frac{2}{N}}
\sum_{j=0}^{N-1}
\vec{e}_j \cos \left[ \frac{2\pi n j}{N} \right]
\,,
\label{eq:cn}
\\
\tilde{\vec{s}}_{n} 
&=
\sqrt{\frac{2}{N}}
\sum_{j=0}^{N-1}
\vec{e}_j \sin \left[ \frac{2\pi n j}{N} \right]
\,.
\label{eq:sn}
\end{align}
Note that $\tilde{\vec{e}}_0$ in \eqref{eq:ezero}
is the basis vector corresponding to the 
$U(1)_{\mathrm{diag}}$,
and it also corresponds to the ``zero-mode"
of the KK modes in the deconstructed dimension.
Notice the $1/\sqrt{N}$ factor in the coefficients,
which leads to the normalization of 
the $U(1)_{\mathrm{diag}}$ gauge coupling \eqref{eq:g4}
in the convention followed in this article.

The inverse transform
for odd $N$ is given as
\begin{align}
\vec{e}_{j} 
=&
\frac{1}{\sqrt{N}}
\left(
{\tilde{\vec{e}}_0}
+
\sum_{n=1}^{\frac{N-1}{2}}
\sqrt{2} \tilde{\vec{c}}_{n}
 \cos \left[\frac{2\pi  n j}{N}\right]
+
\sum_{n=1}^{\frac{N-1}{2}}
\sqrt{2} \tilde{\vec{s}}_{n}
 \sin \left[\frac{2\pi  n j}{N}\right]
\right)
\,.
\label{eq:baseDFTcso}
\end{align}
The inverse transform
for even $N$ is given by
\begin{align}
\vec{e}_{j} 
=&
\frac{1}{\sqrt{N}}
\left(
{\tilde{\vec{e}}_0}
+
{\tilde{\vec{e}}_{\frac{N}{2}}}
(-)^{j}
+
\sum_{n=1}^{\frac{N-2}{2}}
\sqrt{2} \tilde{\vec{c}}_{n}
 \cos \left[\frac{2\pi  n j}{N}\right]
+
\sum_{n=1}^{\frac{N-2}{2}}
\sqrt{2} \tilde{\vec{s}}_{n}
 \sin \left[\frac{2\pi  n j}{N}\right]
\right)
\,.
\label{eq:baseDFTcse}
\end{align}
The mass eigenvalue of the $n$-th KK mode
(for both cosine and sine basis vectors)
is given by
\cite{Hill:2000mu,Arkani-Hamed:2001kyx,Furuuchi:2020klq}
\begin{equation}
m_{A(n)}^2
=
2 g^2 f^2
\left[
1 - \cos \left( \frac{2\pi n}{N}\right)
\right]
=
4 g^2 f^2 \sin^2 \left( \frac{\pi n}{N} \right)
\,.
\label{eq:mAn}
\end{equation}

\section{The Internal Structure of the Dirac Monopole 
in the Low Energy EFT}\label{sec:DiracMonopole}

In this section, I study 
how the inside of the unit magnetic charge Dirac monopole in the Low Energy EFT
is described by the High Energy EFT.
To describe Dirac monopoles,
it is convenient to relabel
the space coordinates as follows:
\begin{equation}
x^1 = x\,,\quad x^2=y\,, \quad x^3=z\,.
\label{eq:xyz}
\end{equation}
It is also convenient to introduce
the radial coordinate
\begin{equation}
r = \sqrt{x^2+y^2+z^2}\,.
\label{eq:r}
\end{equation}
The Dirac monopole solution
with a unit magnetic charge 
in the Low Energy EFT
is given by 
\begin{align}
A_x &= \frac{2\pi}{g_4} \frac{-y}{r(r+z)}\,,
\nn\\
A_y &= \frac{2\pi}{g_4} \frac{x}{r(r+z)}\,,
\nn\\
A_z &= 0\,.
\label{eq:DiracMonopole}
\end{align}

I follow the convention in which 
the smallest electric charge unit is one
with respect to the gauge coupling $g_4$ in the Low Energy EFT.
Thus, the magnetic charge is quantized in the unit
\begin{equation}
g_{4m}:= \frac{2\pi}{g_4} = \frac{2\pi \sqrt{N}}{g}\,,
\label{eq:g4m}
\end{equation}
due to the Dirac quantization condition.

It is straightforward to embed the
Dirac monopole solution \eqref{eq:DiracMonopole} 
in the High Energy EFT
\cite{Poppitz:2003uz}:
\begin{align}
A_{x\, (j)}&= \frac{2\pi}{g} \frac{-y}{r(r+z)}\,,
\nn\\
A_{y\, (j)} &= \frac{2\pi}{g} \frac{x}{r(r+z)}\,,
\nn\\
A_{z\,(j)} &= 0\,, \qquad \qquad \qquad \qquad (\mbox{for all } j)\,. 
\label{eq:DiracMonopoleHigh}
\end{align}


From \eqref{eq:Amuzero} and \eqref{eq:g4},
one observes that 
each Dirac monopole of $U(1)_{(j)}$
contributes to the magnetic charge 
$1/N$ of $U(1)_{\mathrm{ diag}}$ gauge group
in the unit of $g_{4} = 2\pi/g_4$.
Therefore, the total magnetic charge in $U(1)_{\mathrm{ diag}}$ gauge group
from $N$ Dirac monopoles each from $U(1)_{(j)}$ gauge group
adds up to one
(and no net magnetic charge in the broken $U(1)$ gauge groups).

Note that 
the solution has to take the same functional form
for all $j$ ($j=0,1,\cdots,N-1$),
i.e. all the Dirac monopoles of $U(1)_{(j)}$ gauge group
are top of each other ($r=0$ in the case above),
otherwise, the KK modes in the latticized extra dimension
would be excited and cost extra energy.

\begin{figure}[htbp]
\centering
\includegraphics[width=4in]{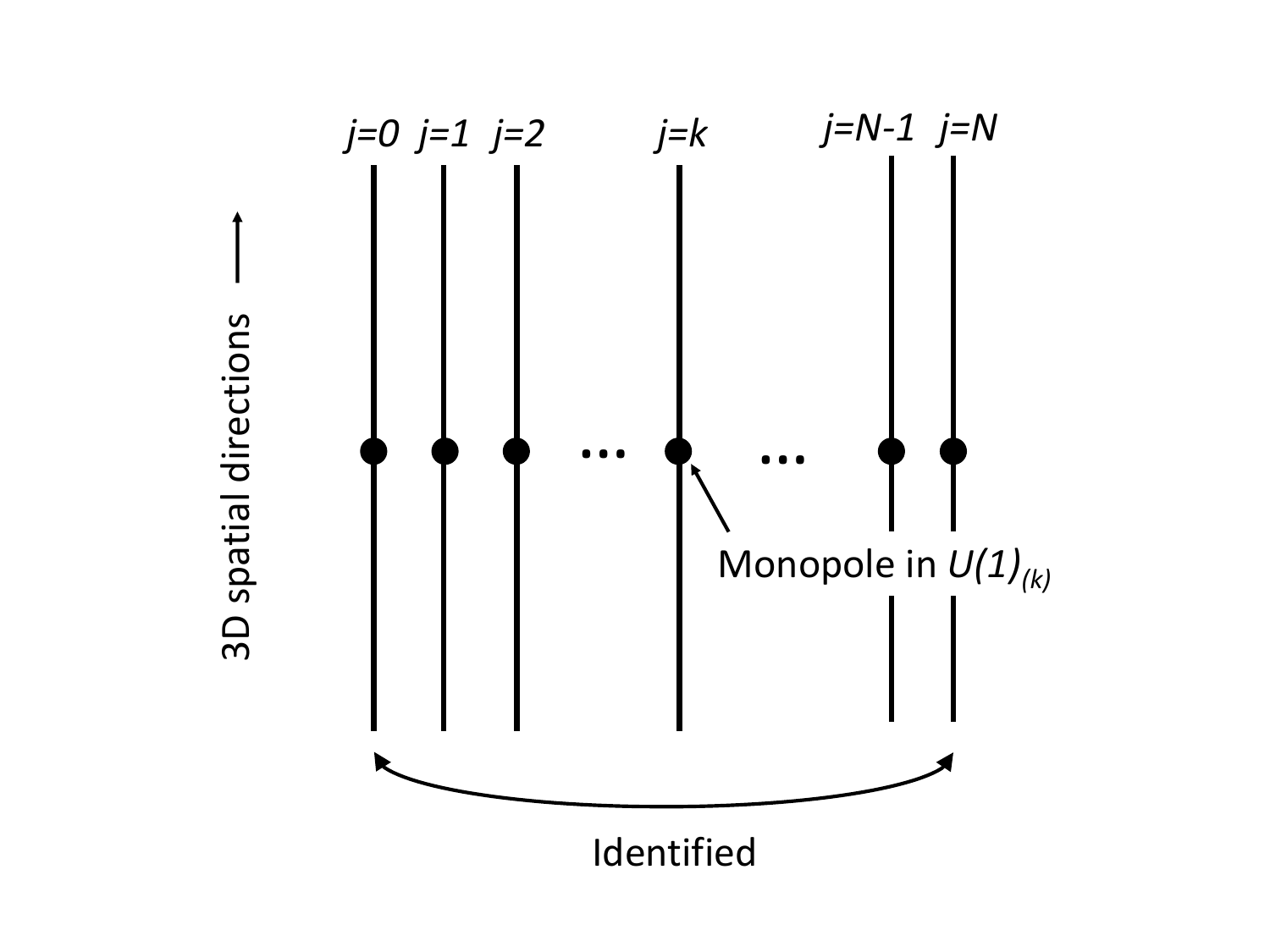} 
\caption{The Dirac monopole in 
the Low Energy EFT can be described
as a monopole loop wrapping around the
latticized extra dimension
in the High Energy EFT.
\label{fig:monopoleloop}}
\end{figure}

The solution is invariant
under the discrete translation $j \rightarrow j+1$ \eqref{eq:dts}.
The solution
\eqref{eq:DiracMonopoleHigh} may be called
``monopole loop''
wrapping around the latticized extra dimension
(see Fig.~\ref{fig:monopoleloop}).
As explained in the introduction,
while \eqref{eq:DiracMonopoleHigh}
is a classical solution of the High Energy EFT,
it is an \emph{unstable} solution.
The reason is that
each Dirac monopole in $U(1)_{(j)}$
has a magnetic charge $1/N$ 
in the $U(1)_{\mathrm{ diag}}$ gauge group,
hence they repel each other
by the magnetic Coulomb repulsive forces.
In the High Energy EFT,
the bound state
of $U(1)_{(j)}$ monopoles
must be stabilized by
attractive forces that counter
the magnetic Coulomb repulsive forces.
I argue below that the attractive forces 
are provided by
the Nielsen-Olesen magnetic type flux tubes
\cite{Nielsen:1973cs}
connecting
the $U(1)_{(j)}$ monopoles \cite{Nambu:1974zg}.
The configuration is qualitatively
similar to the ones studied
in \cite{Saraswat:2016eaz,Furuuchi:2017upe}
in the context of examining the WGC \cite{ArkaniHamed:2006dz} at low energy.
The size of the bound state of $U(1)_{(j)}$ monopoles
is determined by the balance
between the repulsive forces
and the attractive forces. 
I will call
the Dirac monopoles each from each $U(1)_{(j)}$ gauge group $(j=1, \cdots, N-1)$ 
``constituent Dirac monopoles'', and
the Dirac monopole in the Low Energy EFT
described as a bound state of the constituent Dirac monopoles
``composite magnetic monopole'' (CMM).

\subsection{The Dirac Monopole of $U(1)_{\mathrm{ diag}}$ 
as a Bound State of the Constituent Dirac Monopoles}\label{sec:CMM}

Below, I describe how the bound state
of the constituent Dirac monopoles
in $U(1)_{(j)}$ gauge groups
of the High Energy EFT
makes up the Dirac monopole of the Low Energy EFT
in the $U(1)_{\mathrm{diag}}$ group 
(composite magnetic monopole, CMM).
I first study the examples
with small $N$,
in particular,
$N=2$ case allows explicit analysis.
Then,
I make an order of magnitude estimate 
for the size of the CMM in the large $N$ limit. 

\subsubsection*{Example: $N=2$ case}\label{sec:N2}

The mass eigen-state basis is given as
\begin{align}
\tilde{\vec{e}}_{0}
&=
\frac{1}{\sqrt{2}} 
\left(\vec{e}_{0} + \vec{e}_{1} \right)\,,
\label{eq:2te0}\\
\tilde{\vec{e}}_{1}
&=
\frac{1}{\sqrt{2}}
\left(\vec{e}_{0} - \vec{e}_{1} \right)
\,.
\label{eq:2te1}
\end{align}
The inverse relation is given by
\begin{align}
{\vec{e}}_{0}
&=
\frac{1}{\sqrt{2}} 
\left( \tilde{\vec{e}}_{0} + \tilde{\vec{e}}_{1} \right)\,,
\label{eq:2e0}\\
{\vec{e}}_{1}
&=
\frac{1}{\sqrt{2}}
\left( \tilde{\vec{e}}_{0} - \tilde{\vec{e}}_{1} \right)\,.
\label{eq:2e1}
\end{align}

In the case $N=2$, the deconstruction model
reduces to $Z=1$ case of
a slightly extended version of the model by Saraswat \cite{Saraswat:2016eaz},
where $Z$ is the charge of the Higgs field 
in one of two $U(1)$ gauge groups in the model.
A slight difference is that
the original model of \cite{Saraswat:2016eaz} has one Higgs field,
while the dimensional deconstruction with $N=2$ has two Higgs fields
with the same charge, namely, $H_{(0,1)}$ and $H_{(1,0)}^\ast$.
The CMM in the original model by Saraswat has been studied in some detail
in \cite{Saraswat:2016eaz,Furuuchi:2017upe,Pathak:2025ukb}.

It should be noted that if one truncates the deconstruction model to the
$U(1)_{\tilde{\vec{e}}_1}$ gauge group sector,
it is essentially the familiar Abelian-Higgs model
that is used to study magnetic flux tube solutions
\cite{Nielsen:1973cs}
(see, for example, \cite{Weinberg:2012pjx} for a textbook account).
A slight difference from the familiar case is, again,
that the deconstruction model has two Higgs fields with the same charge.

I start with
the Dirac monopole in $U(1)_{(j=0)}$ gauge group
and that in the $U(1)_{(j=1)}$ gauge group,
each has the smallest unit of magnetic charge
in each gauge group.
From the Dirac quantization condition,
their magnetic charge is given by
\begin{equation}
g_m := \frac{2\pi}{g}\,,
\label{eq:N2gm}
\end{equation}
in each gauge group.
%
Then, eq.~\eqref{eq:2e0} 
means that
the unit charge in $U(1)_{(j=0)}$ gauge group
has charges $1/\sqrt{2}$ in the gauge groups
$U(1)_{\tilde{\vec{e}}_{0}}$ and $U(1)_{\tilde{\vec{e}}_{1}}$,
in the unit of the original gauge coupling $g$.
However, I should apply the same normalization convention for the gauge couplings
in $U(1)_{\tilde{\vec{e}}_{0}}$ and $U(1)_{\tilde{\vec{e}}_{1}}$ gauge groups,
i.e. so that the smallest electric charge unit is one in each gauge group.
Therefore, I normalize the gauge couplings as
\begin{equation}
g_4 =
g_{4,\tilde{\vec{e}}_{0}} 
= 
g_{4,\tilde{\vec{e}}_{1}} = \frac{g}{\sqrt{2}}\,.
\label{eq:g4N2}
\end{equation}
Note that $\tilde{\vec{e}}_{0}$ is the basis vector
in the direction of the unbroken
$U(1)_{\mathrm{diag}}$ gauge group, so
this part of \eqref{eq:g4N2} is nothing but
\eqref{eq:g4} applied to the $N=2$ case.


The unit of magnetic charge in $\tilde{\vec{e}}_{j}$ gauge group
is 
given by the smallest electric charge (Dirac quantization condition):
\eqref{eq:g4N2}:
\begin{equation}
g_{4m}
=
g_{4m,\tilde{\vec{e}}_{j}} 
= 
\frac{2\pi}{g_{4,\tilde{\vec{e}}_{j}}}
=
\frac{2\sqrt{2}\pi}{g}\,,
\quad (j=0,1)\,.
\label{eq:tgm}
\end{equation}
As the Dirac monopole with unit magnetic charge in 
$U(1)_{(j=0)}$ gauge group has
$1/\sqrt{2}$ magnetic charge in the original unit \eqref{eq:N2gm},
in the unit $g_{4m,\tilde{\vec{e}}_{1}}$
of the $U(1)_{\tilde{\vec{e}}_{1}}$ gauge group,
it has
$1/2$ magnetic charge.
Similarly,
the Dirac monopole with unit magnetic charge in 
$U(1)_{(j=1)}$ gauge group has
$-1/2$ magnetic charge in the unit $g_{4m}^{\tilde{\vec{e}}_{1}}$.

Now, I turn my attention to the magnetic flux tubes \cite{Nielsen:1973cs}
that connect magnetic monopoles \cite{Nambu:1974zg}.
While the magnetic flux from the monopoles
appears as half of the usual,
this is because I used the Dirac quantization condition
with respect to the smallest electric charge in each $U(1)_{\tilde{\vec{e}}_j}$ gauge groups.
If one uses the Higgs charge as a unit instead,
the solution is the familiar
vorticity one solution 
in the Abelian-Higgs model (see e.g. \cite{Weinberg:2012pjx}).
The only difference is that
there are two Higgs fields with the same charge.
That only changes the functional form of the solution,
which one anyway has to solve numerically.

The tension 
of the magnetic flux tube of the broken $U(1)_{\tilde{\vec{e}}_{1}}$ direction
is estimated to be
\begin{equation}
T_{\tilde{\vec{e}}_{1}}
\simeq
{f^2}
\,.
\label{eq:Tte1}
\end{equation}

\begin{figure}[htbp]
\centering
\includegraphics[width=5in]{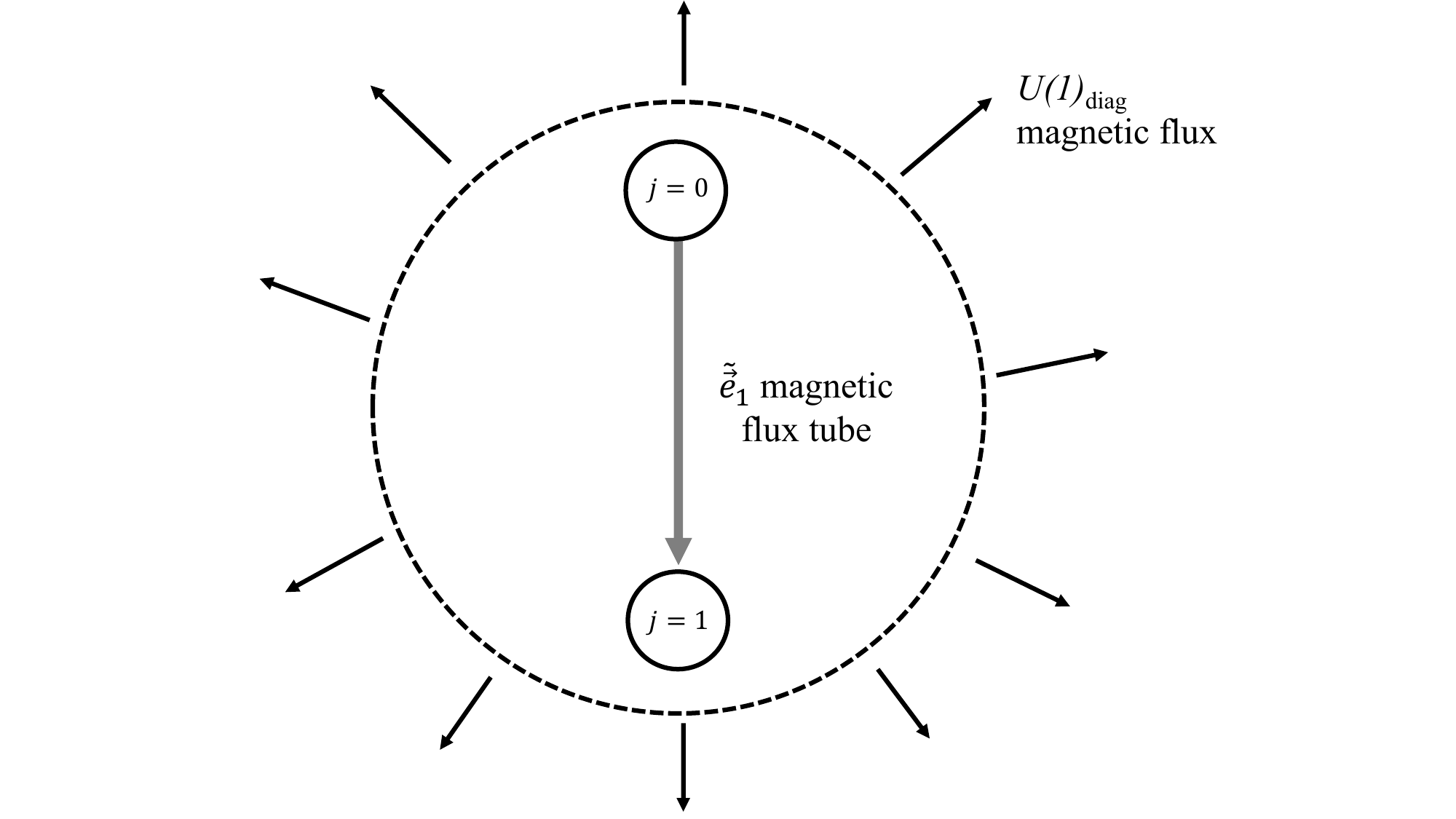} 
\caption{The Dirac monopole in 
the Low Energy EFT can be described
as a bound state of Dirac monopoles in
$U(1)_{(j)}$ ($j=0,1$) gauge group
in the High Energy EFT.
The size of the bound state is 
determined by the balance between the 
repulsive magnetic Coulomb force
and the attractive force due to the
tension of the magnetic flux tube connecting 
the Dirac monopole in the gauge group $U(1)_{(j=0)}$
and that in the gauge group $U(1)_{(j=1)}$.
\label{fig:N2monopole}}
\end{figure}

As mentioned earlier, the size of the CMM,
$L_{\mathrm{CMM}}$, 
is determined by the balance between
the magnetic Coulomb repulsive force of the unbroken gauge group
and
the attractive force due 
to the tension of the magnetic flux tube of the broken gauge group.
The Dirac monopole of $U(1)_{\mathrm{diag}}$ 
with unit magnetic charge
is made of constituent Dirac monopoles
and the magnetic flux tube between them,
see Fig.~\ref{fig:N2monopole}.
\begin{figure}[htbp]
\centering
\includegraphics[width=5in]{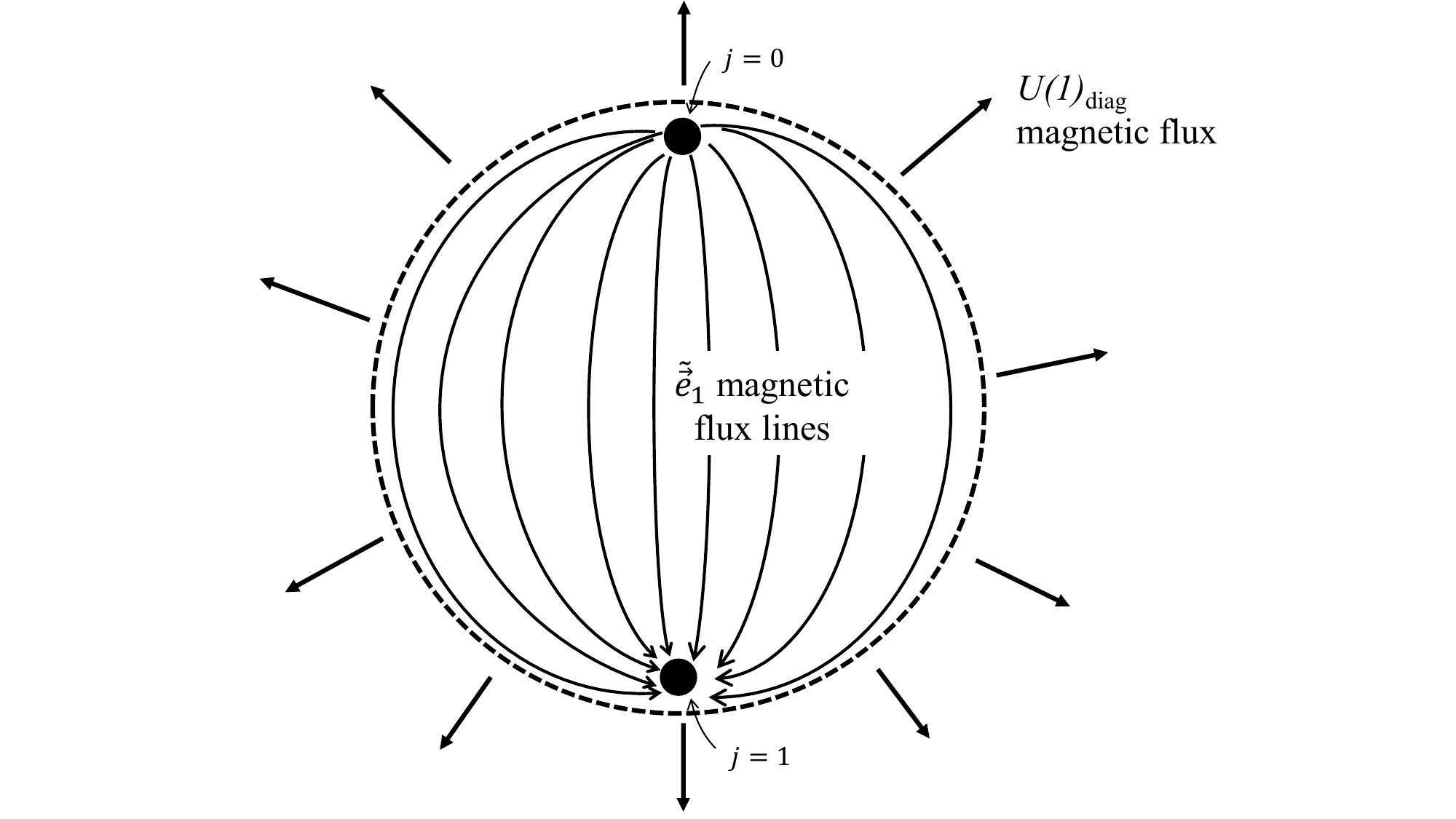} 
\caption{The internal structure of the CMM for the case $N=2$
slightly more elaborated than Fig.~\ref{fig:N2monopole}
in that the radius of the magnetic flux tube is taken into account.
Both the radius and the length of the magnetic flux tube
are of the order of $1/(gf)$,
justifying a posteriori the estimate of the size of the CMM,
regarding the region inside the CMM as approximately spherical.
\label{fig:N2monopoleActual}}
\end{figure}

In terms of the potential, 
the force balance can be written as
\begin{equation}
g_{4m}
\frac{1}{L_{\mathrm{CMM}}}
\simeq
f^2
L_{\mathrm{CMM}}
\,,
\label{eq:N2balance}
\end{equation}
from which we obtain
\begin{equation}
L_{\mathrm{CMM}} 
\simeq
\frac{1}{gf}
\,.
\label{eq:N2Llow}
\end{equation}
In the above estimate,
I was not concerned about numerical factors of order one.

In the above, I neglected the fact that
the radius of the magnetic flux tube
is of the order of $1/(g_4 f) \sim 1/(gf)$
(see Appendix~\ref{sec:aN2}).
The above estimate of the size of the CMM is justified
a posteriori,
regarding the internal configuration as approximately spherical,
with the radial size $L_{\mathrm{CMM}}$.
This does not alter the order of magnitude estimate
of the energy stored inside the CMM.


\subsubsection*{Example: $N=3$ case}\label{sec:N3}

In the examples below,
I only illustrate the flow of the magnetic flux 
between the constituent Dirac monopoles through the magnetic flux tubes,
as the explicit construction of the magnetic flux tube configurations
is not very illuminating.

\begin{figure}[htbp]
\centering
\includegraphics[width=5in]{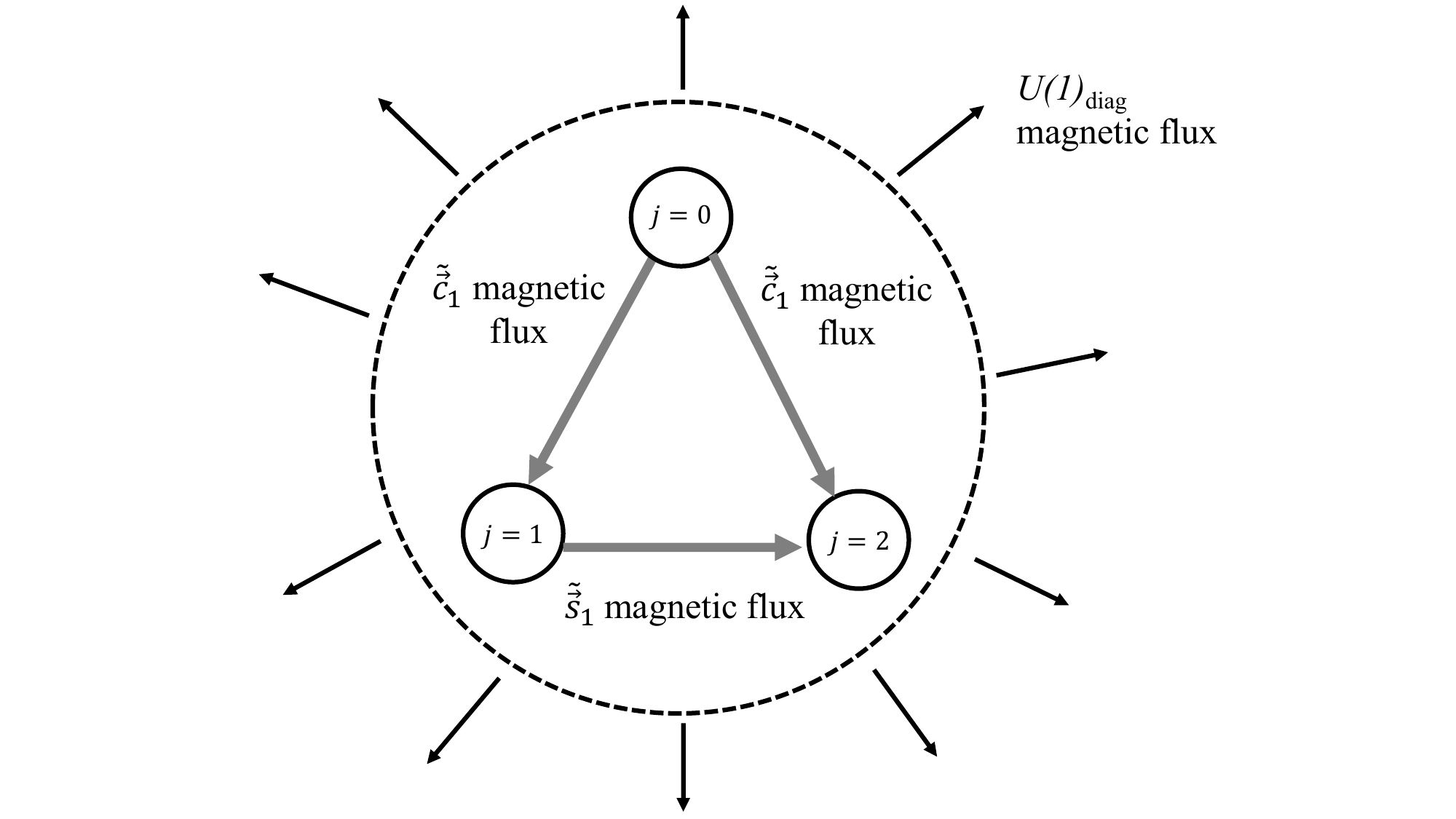} 
\caption{The flow of the magnetic flux inside the CMM for $N=3$.
\label{fig:N3monopole}}
\end{figure}

The ortho-normal basis of the mass eigen-states is given by
\begin{align}
\tilde{\vec{e}}_{0}
&=
\frac{1}{\sqrt{3}} 
\left(\vec{e}_{0} + \vec{e}_{1} + \vec{e}_{2}\right)\,,
\label{eq:3te0}\\
\tilde{\vec{c}}_{1}
&=
\sqrt{\frac{2}{3}}
\vec{e}_{0} 
-\frac{1}{\sqrt{6}} \vec{e}_{1} 
-\frac{1}{\sqrt{6}} \vec{e}_{2} 
\,,
\label{eq:3c1}
\\
\tilde{\vec{s}}_{1}
&=
\frac{1}{\sqrt{2}} 
\vec{e}_{1} 
- 
\frac{1}{\sqrt{2}} \vec{e}_{2} 
\,.
\label{eq:3s1}
\end{align}
The inverse relations are given by
\begin{align}
\vec{e}_{0}
=&
\frac{1}{\sqrt{3}} 
\tilde{\vec{e}}_{0}
+ 
\frac{2}{\sqrt{6}}
\tilde{\vec{c}}_{1}\,,
\label{eq:3e0}\\
\vec{e}_{1} 
=&
\frac{1}{\sqrt{3}}
\tilde{\vec{e}}_{0}
-
\frac{1}{\sqrt{6}} 
\tilde{\vec{c}}_{1}
+
\frac{1}{\sqrt{2}} 
\tilde{\vec{s}}_{1}
\,,
\label{eq:3e1}\\
\vec{e}_{2} 
=&
\frac{1}{\sqrt{3}}
\tilde{\vec{e}}_{0}
-
\frac{1}{\sqrt{6}} \tilde{\vec{c}}_{1}
-
\frac{1}{\sqrt{2}} \tilde{\vec{s}}_{1}
\,.
\label{eq:3e2}
\end{align}
The flow of the magnetic flux
through the magnetic flux tubes
can be read off from 
\eqref{eq:3e0}-\eqref{eq:3e2},
and schematically depicted in Fig.~\ref{fig:N3monopole}.

\subsubsection*{Example: $N=4$ case}\label{sec:N4}

The ortho-normal basis of mass eigen-states is given as
\begin{align}
\tilde{\vec{e}}_{0}
&=
\frac{1}{2} 
\left(
\vec{e}_{0} + \vec{e}_{1} + \vec{e}_{2} + \vec{e}_{3} 
\right)\,,
\label{eq:4te0}\\
\tilde{\vec{e}}_{2}
&=
\frac{1}{2} 
\left(
\vec{e}_{0} - \vec{e}_{1} + \vec{e}_{2} - \vec{e}_{3} 
\right)\,,
\label{eq:4te2}\\
\tilde{\vec{c}}_{1}
&=
\frac{1}{\sqrt{2}}
\left(
\vec{e}_{0} - \vec{e}_{2} 
\right)
\,,
\label{eq:4c1}
\\
\tilde{\vec{s}}_{1}
&=
\frac{1}{\sqrt{2}}
\left( 
\vec{e}_{1} -  \vec{e}_{3} 
\right)
\,.
\label{eq:4s1}
\end{align}
The inverse relations are given as
\begin{align}
\vec{e}_{0}
=&
\frac{1}{2} \tilde{\vec{e}}_{0}
+ 
\frac{1}{2} \tilde{\vec{e}}_{2}
+ 
\frac{1}{\sqrt{2}} 
\tilde{\vec{c}}_{1} \,,
\label{eq:4e0}\\
\vec{e}_{1} 
=&
\frac{1}{2}
\tilde{\vec{e}}_{0}
-
\frac{1}{2} 
\tilde{\vec{e}}_{2}
+
\frac{1}{\sqrt{2}} 
\tilde{\vec{s}}_{1}
\,,
\label{eq:4e1}\\
\vec{e}_{2} 
=&
\frac{1}{2}
\tilde{\vec{e}}_{0}
+ 
\frac{1}{2} 
\tilde{\vec{e}}_{2}
-
\frac{1}{\sqrt{2}} 
\tilde{\vec{c}}_{1}
\,,
\label{eq:4e2}\\
\vec{e}_{3} 
=&
\frac{1}{2}
\tilde{\vec{e}}_{0}
-
\frac{1}{2} 
\tilde{\vec{e}}_{2}
-
\frac{1}{\sqrt{2}} 
\tilde{\vec{s}}_{1}
\,.
\label{eq:4e3}
\end{align}

\begin{figure}[htbp]
\centering
\includegraphics[width=5in]{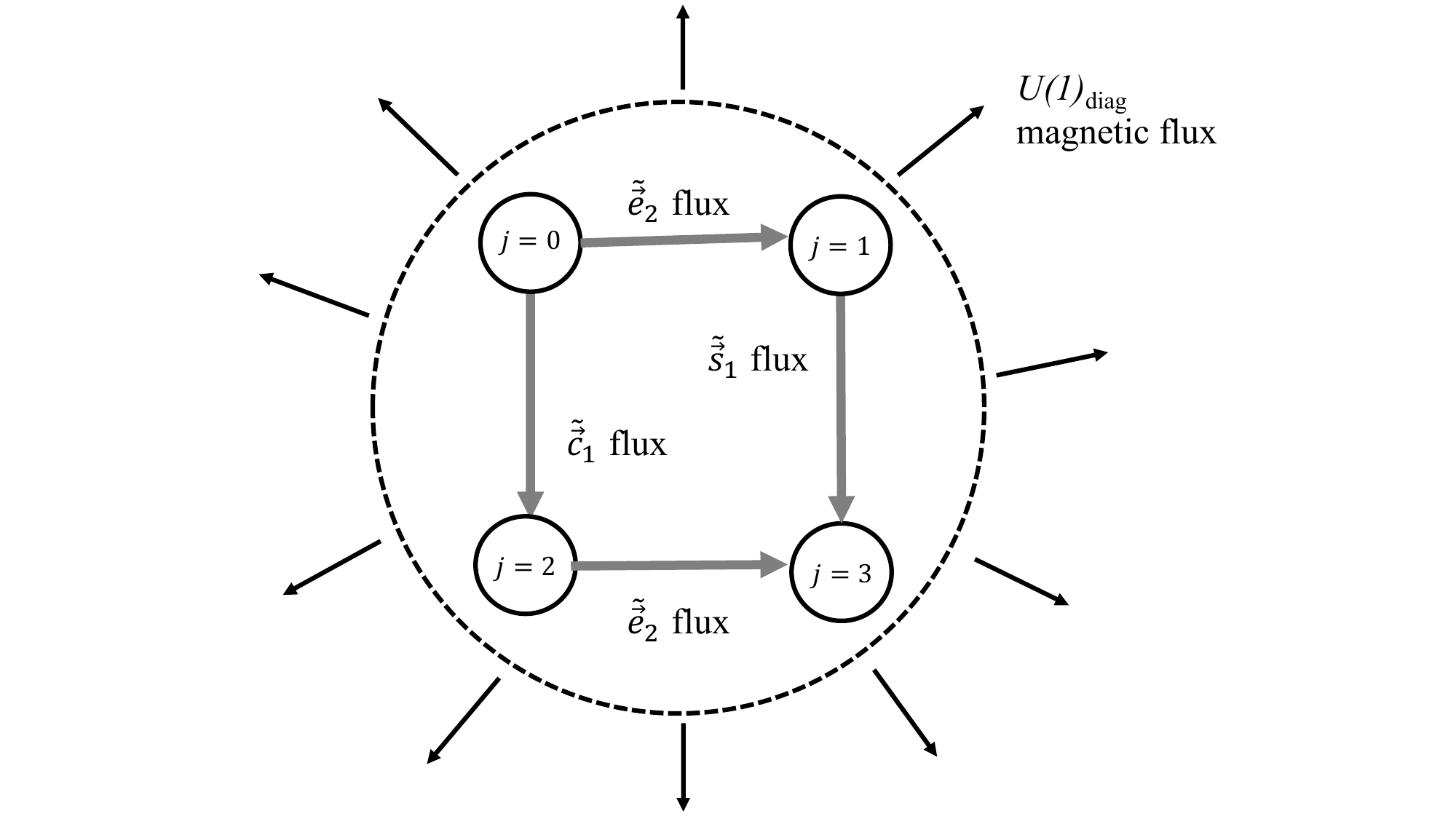} 
\caption{The flow of the magnetic flux inside the CMM for $N=4$.
\label{fig:N4monopole}}
\end{figure}

The flow of the magnetic flux
through the magnetic flux tubes
can be read off from 
\eqref{eq:4e0}-\eqref{eq:4e3},
and schematically  depicted in Fig.~\ref{fig:N4monopole}.

\subsubsection*{The Size of the $U(1)_{\mathrm{diag}}$ Monopole at Large $N$}\label{sec:largeN}

Now I estimate the
size of the $U(1)_{\mathrm{diag}}$ monopole at large $N$.
For this purpose,
the previous examples with small $N$
were somewhat deceptive,
in that the charge ratios in the mass eigen-state basis are generically irrational for larger $N$.
To estimate the size of the CMM qualitatively,
I consider magnetic flux tubes
in $U(1)_{(j)}$ gauge groups
(see Appendix~\ref{sec:avk})
as an approximation.

\begin{figure}[htbp]
\centering
\includegraphics[width=5in]{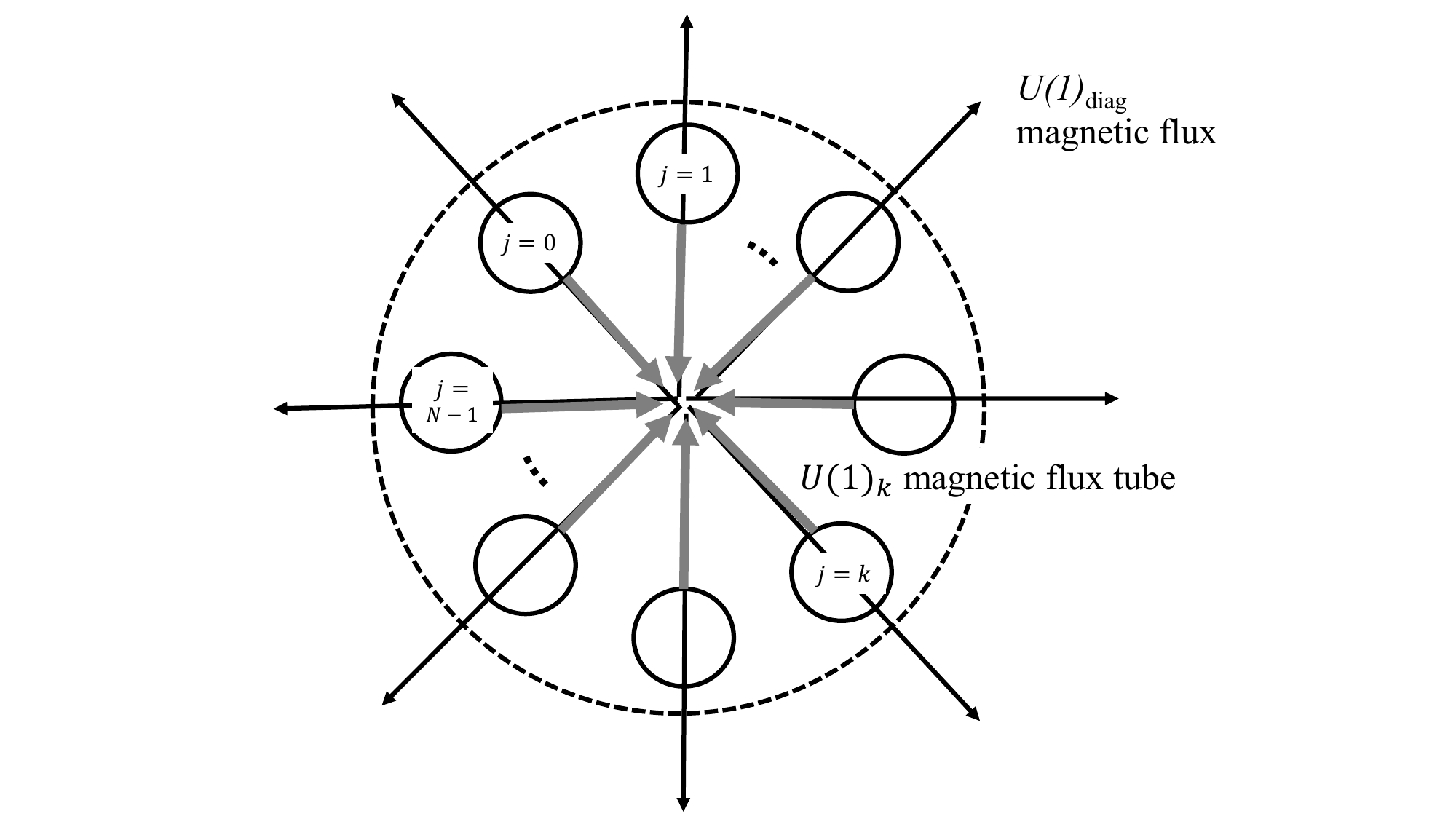} 
\caption{A schematic figure for estimating the 
energy and the size of the
$U(1)_{\mathrm{diag}}$ monopople.
From each Dirac monopole in $U(1)_{(j)}$ gauge group,
$U(1)_{(j)}$ magnetic flux tube (shown in the grey ingoing arrow) carries out
the $2\pi/g$ amount of the magnetic flux of $U(1)_{(j)}$ gauge group.
However, $2\pi/(g\sqrt{N}) = 2\pi/(g_4 N)$ amount of $U(1)_{\mathrm{diag}}$ magnetic flux
should not be confined to the $U(1)_{(j)}$ magnetic flux tube;
that is cancelled with the outgoing $U(1)_{\mathrm{diag}}$ magnetic flux (shown in the black outgoing arrow).
The rest of the magnetic fluxes in the magnetic flux tubes cancel at the ``centre'':
note that this schematic configuration is not meant to be an exact configuration,
rather, it is just to estimate the approximate size of the bound state.
\label{fig:LargeNmonopole}}
\end{figure}

The size of the bound state 
is determined by the
balance between the magnetic Coulomb repulsive potential
due to the unbroken gauge group $U(1)_{\mathrm{dia}g}$
and that of the tension of the magnetic flux tubes.
I estimate the energy from the magnetic flux tubes as follows:
From each Dirac monopole in $U(1)_{(j)}$ gauge group,
$U(1)_{(j)}$ magnetic flux tube 
(shown in the grey ingoing arrow in Fig.~\ref{fig:LargeNmonopole}) 
carries out
the $g_m = 2\pi/g$ amount of the magnetic flux of $U(1)_{(j)}$ gauge group.
The tension of the magnetic flux tube is of the order of $f^2$, see Appendix~\ref{sec:avk}.
However, $2\pi/(g\sqrt{N}) = 2\pi/(g_4 N) = g_{4m}/N$ amount of the $U(1)_{\mathrm{diag}}$ magnetic flux
should not be confined to the $U(1)_{(j)}$ magnetic flux tube;
that is cancelled with the outgoing $U(1)_{\mathrm{diag}}$ magnetic flux 
(shown in the black outgoing arrow in Fig.~\ref{fig:LargeNmonopole}).
Note that none of the Higgs fields is charged under the $U(1)_{\mathrm{diag}}$ gauge group,
so $U(1)_{\mathrm{diag}}$ magnetic flux are not squeezed.
However, in the large $N$ limit, the density of the magnetic flux tubes becomes high,
and assuming the outgoing $U(1)_{\mathrm{diag}}$ magnetic flux to cancel in the $U(1)_{(j)}$ magnetic flux tubes
would not change the order of magnitude estimate of the energy stored in the $U(1)_{(j)}$ magnetic flux tubes.
The rest of the magnetic fluxes cancel at the centre of the CMM.
One may regard the $U(1)_{(j)}$ magnetic flux tube
as all the magnetic fluxes in the mass eigen-state basis squeezed into a single tube
(compare with Fig.~\ref{fig:N3monopole}. \ref{fig:N4monopole}),
with extra $U(1)_{\mathrm{diag}}$ magnetic flux in addition.

From the above considerations,
the balance between the repulsive magnetic Coulomb potential
of unbroken $U(1)_{\mathrm{diag}}$ gauge group
and 
the attractive potential from the tension of the magnetic flux tubes
are achieved when
\begin{equation}
\frac{N(N-1)}{2}
\left( \frac{g_{4m}}{{N}} \right)^2
\frac{1}{L_{\mathrm{CMM}}}
\simeq
N f^2
L_{\mathrm{CMM}}\,.
\label{eq:Nbalance}
\end{equation}
%
%
From \eqref{eq:Nbalance}, 
I obtain the size of the CMM as
\begin{equation}
L_{\mathrm{CMM}}
\simeq \frac{1}{gf}
\simeq d \,,
\label{eq:Llow}
\end{equation}
where $d$ is the lattice spacing defined in \eqref{eq:d}.
Here, I suppressed numerical factors of order one.

As in the case for $N=2$,
Fig.~\ref{fig:LargeNmonopole}
only captures the flow of the magnetic fluxes.
If one considers the radial size of the magnetic fluxes $\sim 1/(gf)$,
the fluxes are smeared out in the volume with radius $L_{\mathrm{CMM}} \sim 1/(gf)$,
like in Fig.~\ref{fig:N2monopoleActual}. 
This smearing does not alter the order of magnitude estimate of the energy
stored in the region.

The result \eqref{eq:Llow} is understandable: 
If one regards the deconstructed theory
as 5D theory with UV cut-off scale $1/d$,
the deviation from the continuum Lorentz-invariant
5D theory shows up at this scale.
Note that
in the continuum 5D scace time,
the monopole loop solution,
i.e. 4D Dirac magnetic monopoles
placed in a translationally invariant fashion in the 5th-dimensional circle,
is a stable solution (the continuum limit of Fig.~\ref{fig:monopoleloop}),
as mentioned in the introduction section.

\section{Briefly on the WGC}\label{sec:WGC}

The Weak Gravity Conjecture (WGC) \cite{ArkaniHamed:2006dz}
is a proposed consistency criterion for 
Abelian gauge theories coupled to quantum gravity.
For a $U(1)$ gauge theory in 4D to be consistently coupled to quantum gravity,
the WGC mandate the existence of a state whose charge $q$ and mass $m$ satisfy
(electric WGC)
\begin{equation}
g q\gtrsim \frac{m}{M_P} \,,
\label{eq:eWGC}
\end{equation}
where $g$ is the gauge coupling and $M_P$ is the reduced Planck mass.

Applying a similar condition for
magnetic charges (magnetic WGC),
%
the following upper bound on the 
UV cut-off scale of the EFT is obtained:
\begin{equation}
\Lambda_{\mathrm{UV}} \lesssim g M_P \, .
\label{eq:cutWGC}
\end{equation}
In deriving \eqref{eq:cutWGC},
one uses the lower bound on the mass $m_{\mathrm{mono}}$ of the Dirac monopole
estimated from EFT:
\begin{equation}
m_{\mathrm{mono}} \gtrsim \frac{\Lambda_{\mathrm{UV}}}{g^2} \,.
\label{eq:mmono}
\end{equation}

A generalization of the WGC to product $U(1)$ gauge groups
has been studied in \cite{Cheung:2014vva}.
In the current work,
the High Energy EFT has the product
$\prod_{j=0}^{N-1} U(1)_{(j)}$ gauge group,
and the charges and the masses of the fields
are the same for all the $U(1)_{(j)}$ groups
due to the discrete translational symmetry
\eqref{eq:dts}.
In this case,
the electric WGC asserts
\begin{equation}
\frac{g }{\sqrt{N} }
\gtrsim \frac{m}{M_P} \,.
\label{eq:HEeWGC}
\end{equation}
The bound on the UV cut-off scale from the magnetic WGC becomes
\begin{equation}
\Lambda_{\mathrm{UV}} \lesssim \frac{g M_P}{\sqrt{N}} \, .
\label{eq:cutWGCN}
\end{equation}
Here, one may consider $\Lambda_{\mathrm{UV}}$ 
as the UV cut-off scale of the
High Energy EFT \eqref{eq:S}.
However, since the gauge coupling $g_4$
of the Low Energy EFT is related to 
the gauge coupling $g$ of the High Energy EFT as in \eqref{eq:g4}:
\begin{equation}
g_4 = \frac{g}{\sqrt{N}} \,,
\end{equation}
and the UV cut-off scale of the Low Energy EFT
should be lower than that of the High Energy EFT,
it follows from \eqref{eq:cutWGCN} that
\begin{equation}
\Lambda_{\mathrm{low}} < \Lambda_{\mathrm{UV}} \lesssim g_4 M_P\, .
\label{eq:cutWGCNlow}
\end{equation}
Eq.~\eqref{eq:cutWGCNlow} means that
if one assumes that the High Energy EFT satisfies
the magnetic WGC bound \eqref{eq:cutWGCN}
for the product $U(1)^N$ gauge theory,
the magnetic WGC bound for the Low Energy EFT automatically follows from the assumption.

In fact, one could instead assume that the dimensional deconstruction EFT
\eqref{eq:Sdec} satisfies the magnetic WGC bound:
\begin{equation}
\Lambda_{\mathrm{pert}} \lesssim \frac{g M_P}{\sqrt{N}} \,.
\label{eq:cutWGCNdec}
\end{equation}
Again, it directly follows from the assumption \eqref{eq:cutWGCNdec} 
that the Low Energy EFT satisfies the magnetic WGC bound.

The situation is quite different from the model due to Saraswat
\cite{Saraswat:2016eaz,Furuuchi:2017upe}:
there, the size of the CMM signified the
breakdown of the Low Energy EFT.
Here, in the case of dimensional deconstruction, 
the Low Energy EFT
\eqref{eq:S4}
breaks down at the KK energy scale, i.e. 
the inverse of the KK compactification radius \eqref{eq:LKK},
which is far below the inverse size of the CMM $\sim$ inverse of the lattice spacing $d$
in the large $N$ limit:
\begin{equation}
\Lambda_{\mathrm{low}} = \frac{1}{L_{\mathrm{KK}}}
=
\frac{gf}{N} 
\ll
gf
=
\frac{1}{d}
\sim
\frac{1}{L_{\mathrm{CMM}}}
\,.
\end{equation}

\section{Summary and Discussions}\label{sec:discussions}

In this article,
I showed that 
the Dirac monopole in a deconstructed gauge theory
has an interesting internal structure
reminiscent of the composite magnetic monopole (CMM)
in the model proposed by Saraswat \cite{Saraswat:2016eaz}.
The size of the CMM was determined by the balance
between the repulsive magnetic Coulomb potential 
of the unbroken $U(1)_{\mathrm{diag}}$ gauge group
and the attractive tension of the magnetic flux tubes
of the broken gauge groups.
I estimated the size of the CMM to be
of the order of the lattice spacing
of the deconstructed dimension.

I started from the perturbative UV model of
dimensional deconstruction based on the Higgs mechanism \cite{Hill:2000mu}
to study the internal structure of the CMM.
This approach had an advantage that
the short-distance internal structure is described classically.
On the other hand, there is another UV model for dimensional deconstruction
based on quiver gauge theories
and chiral symmetry breaking \cite{Arkani-Hamed:2001kyx},
in which pion-like composite fields play the role of the link field for the gauge field
in the latticized direction.
This UV model enters non-perturbative regime at the chiral symmetry breaking scale.
In this formulation,
the radial component of the Higgs field is absent.
This poses an obstruction to
construct 
Nielsen-Olesen type
magnetic flux tubes,
since it is crucial in this type of solution
that the value of the Higgs field becomes zero
at the core of the magnetic flux tube. 
However, within the dimensional deconstruction EFT \eqref{eq:Sdec},
one may construct a magnetic flux tube solution,
expecting that
the behaviour of the solution
would be modified inside the magnetic flux tube
at the distance shorter than the EFT UV cut-off scale.
The tension of the magnetic flux tubes
would be bounded by the UV cut-off scale in this case,
similar to the estimate of the mass of the Dirac monopoles in EFT
\eqref{eq:mmono}.
While it was advantageous to have a perturbative description
in the UV model 
based on the Higgs mechanism \cite{Hill:2000mu},
it will be interesting to study different UV models like \cite{ArkaniHamed:2001nc}.
In this context, it is worthwhile to note that 
there is a swampland conjecture
which proposes that
a radial partner of an axionic field
must appear below a certain energy scale 
related to the Planck scale,
similar to the bound on the UV cut-off scale from the magnetic WGC
\cite{Reece:2018zvv}.

I also investigated
the role of the internal structure of the 
deconstructed Dirac monopole
in the WGC.
Unlike the CMM in the model due to Saraswat
\cite{Saraswat:2016eaz,Furuuchi:2017upe},
however,
it did not have a significant implication
to the WGC in the dimensional deconstruction model.

\vskip8mm
\begin{center}
\textsl{Acknowledgments}
\end{center}
\vskip-1.5mm
I would like to thank Toshifumi Noumi for the hospitality 
during my visit to the University of Tokyo, Komaba, and for the useful discussions.
I also thank Rashmikanta Mantry and  Moreshwar Pathak for helpful discussions. 
This work is supported in part by the project file no.~DST/INT/JSPS/P-344/2022.
Manipal Centre for Natural Sciences, \textsl{Centre of Excellence}, 
Manipal Academy of Higher Education (MAHE) 
is acknowledged for facilities and support.

\appendix
\section{Discrete Fourier Transform}\label{sec:aDFT}

Consider a cyclically ordered $N$ points
labelled by $j$ 
($j = 0, 1, \cdots , N-1$ $(\mathrm{mod}\,\, N)$).
Consider 
a variable $f_j$ which has a value on each point.
I use the following convention for
the discrete Fourier expansion of the variable
$f_j$:
\begin{equation}
f_{j} 
= 
\frac{1}{\sqrt{N}}
\sum_{n=-[\frac{N-1}{2}]}^{[\frac{N}{2}]}
\tilde{f}_n\, e^{i\frac{2\pi  n j}{N}}
\,.
\label{eq:fn}
\end{equation}
Here, $[w]$ denotes the largest integer 
which does not exceed
the real number $w$.
The normalization convention is such that
when applied to dimensional deconstruction
each KK mode is canonically normalized.

The
orthogonality 
of the exponential function:
\begin{equation}
\sum_{j=0}^{N-1}
\left(
e^{i\frac{2\pi  n_1 j}{N}} 
\right)^\ast
e^{i\frac{2\pi  n_2 j}{N}}
= N \delta_{n_1 n_2}\, ,
\label{eq:DFTo}
\end{equation}
leads to the following formula for the
discrete Fourier coefficient:
\begin{equation}
\tilde{f}_n
=
\frac{1}{\sqrt{N}}
\sum_{j=0}^{N-1}
f_j e^{-i\frac{2\pi  n j}{N}}\,.
\label{eq:invDFT}
\end{equation}

The discrete Fourier expansion 
can also be written in terms of $\sin$ and $\cos$ functions.
For odd $N$,
\begin{align}
f_{j} 
=&
\frac{1}{\sqrt{N}}
\left(
{\tilde{f}_0}
+
\sum_{n=1}^{\frac{N-1}{2}}
\left(
\tilde{f}_n
+
\tilde{f}_{-n}
\right)
 \cos \left[\frac{2\pi  n j}{N}\right]
+
\sum_{n=1}^{\frac{N-1}{2}}
i
\left(
\tilde{f}_n
-
\tilde{f}_{-n}
\right)
 \sin \left[\frac{2\pi  n j}{N}\right]
\right)
\nn\\
:=&
\frac{1}{\sqrt{N}}
\left(
{\tilde{f}_0}
+
\sum_{n=1}^{\frac{N-1}{2}}
\sqrt{2} \tilde{f}_{c\,n}
 \cos \left[\frac{2\pi  n j}{N}\right]
+
\sum_{n=1}^{\frac{N-1}{2}}
\sqrt{2} \tilde{f}_{s\,n}
 \sin \left[\frac{2\pi  n j}{N}\right]
\right)
\,.
\label{eq:DFTcso}
\end{align}
For even $N$,\footnote{%
The term proportional to $(-)^{j}$ 
can be included in the sum of cosine functions,
but the coefficient of this term
must be different from the other terms in the sum
in order to make the mass-eigenstate basis vectors ortho-normal.}
\begin{align}
f_{j} 
=&
\frac{1}{\sqrt{N}}
\left(
{\tilde{f}_0}
+
{\tilde{f}_{\frac{N}{2}}}
(-)^{j}
+
\sum_{n=1}^{\frac{N-2}{2}}
\left(
\tilde{f}_n
+
\tilde{f}_{-n}
\right)
 \cos \left[\frac{2\pi  n j}{N}\right]
+
\sum_{n=1}^{\frac{N-2}{2}}
i
\left(
\tilde{f}_n
-
\tilde{f}_{-n}
\right)
 \sin \left[\frac{2\pi  n j}{N}\right]
\right)
\nn\\
:=&
\frac{1}{\sqrt{N}}
\left(
{\tilde{f}_0}
+
{\tilde{f}_{\frac{N}{2}}}
(-)^{j}
+
\sum_{n=1}^{\frac{N-2}{2}}
\sqrt{2} \tilde{f}_{c\,n}
 \cos \left[\frac{2\pi  n j}{N}\right]
+
\sum_{n=1}^{\frac{N-2}{2}}
\sqrt{2} \tilde{f}_{s\,n}
 \sin \left[\frac{2\pi  n j}{N}\right]
\right)
\,.
\label{eq:DFTcse}
\end{align}
When $f_j$ is a real variable,
$\tilde{f}_{-n}^\ast = \tilde{f}_n$
and one obtains
\begin{align}
\tilde{f}_{0}
&=
\frac{1}{\sqrt{N}}
\sum_{j=0}^{N-1}
f_{j}\,,
\label{eq:fzero}\\
\tilde{f}_{\frac{N}{2}}
&=
\frac{1}{\sqrt{N}}
\sum_{j=0}^{N-1}
(-)^j f_{j}\,,
\quad \mbox{(when $N$ is even)}\,,
\label{eq:fNhalf}\\
\tilde{f}_{c\,n} 
&= \sqrt{2}\, \mathrm{Re}\, \tilde{f}_n
= 
\sqrt{\frac{2}{N}}
\sum_{j=0}^{N-1}
f_j \cos \left[ \frac{2\pi n j}{N} \right]
\,,
\label{eq:fcn}
\\
\tilde{f}_{s\,n} &= -\sqrt{2}\, \mathrm{Im}\, \tilde{f}_n
=
\sqrt{\frac{2}{N}}
\sum_{j=0}^{N-1}
f_j \sin \left[ \frac{2\pi n j}{N} \right]
\,.
\label{eq:fcs}
\end{align}

\section{More Details of the $N=2$ Case}\label{sec:aN2}
In the case of $N=2$,
the action \eqref{eq:S} reduces to
\begin{align}
S
=
\int 
 d^4x
\Biggl[
- & \frac{1}{4}
F_{\mu\nu\,(0)}
F^{\mu\nu}_{(0)}
-\frac{1}{4}
F_{\mu\nu\,(1)}
F^{\mu\nu}_{(1)}
+
\frac{1}{2}
D_\mu H_{(0,1)}^\ast
D^\mu H_{(0,1)}
+
\frac{1}{2}
D_\mu H_{(1,0)}^\ast
D^\mu H_{(1,0)}
\nn \\
&
- V(H_{(0,1)})
- V(H_{(1,0)})
+\cdots
\Biggr]\,,
\label{eq:aS}
\end{align}
where the dots denote the action for other charged matter fields.
I assume that the existence of a unit charge field in each $U(1)_{(j)}$ 
with respect to the gauge coupling $g$
(subject to the discrete translational symmetry \eqref{eq:dts}).

The covariant derivatives for the Higgs fields are given as
\begin{align}
D_\mu H_{(0,1)}
&=
\pa_\mu H_{(0,1)} 
- i g A_{\mu (0)} H_{(0,1)} + i g H_{(0,1)} A_{\mu (1)} \,,
\label{eq:aDH01}\\
D_\mu H_{(1,0)}
&=
\pa_\mu H_{(1,0)} 
- i g A_{\mu (1)} H_{(1,0)} + i g H_{(1,0)} A_{\mu (0)} \,.
\label{eq:aDH10}
\end{align}

Moving to
the mass eigen-state ortho-normal basis:
\begin{equation}
A_{\mu,\tilde{\vec{e}}_0}
:=
\frac{1}{\sqrt{2}}
\left(
A_{\mu,(0)} + A_{\mu,(1)}
\right)\,,
\quad
A_{\mu,\tilde{\vec{e}}_1}
:=
\frac{1}{\sqrt{2}}
\left(
A_{\mu,(0)} - A_{\mu,(1)}
\right)\,,
\label{eq:atilde}
\end{equation}
the covariant derivatives
\eqref{eq:aDH01}, \eqref{eq:aDH10} become
\begin{align}
D_\mu H_{(0,1)}
&=
\pa_\mu H_{(0,1)} 
- i 2{g}_4 A_{\mu, \tilde{\vec{e}}_!} H_{(0,1)} \,,
\label{eq:aDH01t}\\
D_\mu H_{(1,0)}
&=
\pa_\mu H_{(1,0)} 
+ i 2 {g}_4 A_{\mu, \tilde{\vec{e}}_1} H_{(1,0)} \,,
\label{eq:aDH10t}
\end{align}
where
\begin{equation}
{g}_4 := \frac{g}{\sqrt{2}}\,,
\label{eq:atildeg}
\end{equation}
was defined in \eqref{eq:g4} for general $N$,
applied here for $N=2$.

A magnetic flux tube solution along $x^3$ axis
may be constructed with the ansatz
(I closely follow the notation of \cite{Weinberg:2012pjx}
with adequate modifications):
\begin{align}
H_{(0,1)}^\ast(x) = H_{(1,0)}(x) &= {f e^{i \theta}}\phi(2{g}_4f \rho)\,,
\label{eq:aFTH}\\
A_{\ell, \tilde{\vec{e}}_1} (x)&= 
\epsilon_{\ell s} \hat{x}^s \frac{a(2{g}_4 f \rho)}{2{g}_4\rho} \,,
\label{eq:aFTA}
\end{align}
where $\ell,s = 1,2$, $\rho := \sqrt{(x^1)^2+ (x^2)^2}$ is
the radial coordinate on the $1$-$2$ plane,
$x^1 = \rho \cos \theta$ and $x^2 = \rho \sin \theta$,
$\hat{x}^s := x^s/\rho$, and
$a$ and $\phi$ are functions to be determined by the equations of motion
with the boundary conditions:
\begin{align}
a(0) =0\,,&\quad \phi(0) = 0\,,
\\
a(\infty) =1\,,&\quad \phi(\infty) = 1\,.
\end{align}
With the anzats \eqref{eq:aFTA}, the magnetic flux is calculated as
\begin{equation}
F_{12,\,\tilde{\vec{e}}_1}
=
-\frac{1}{2g_4 \rho}
\frac{\pa a}{\pa \rho} \,.
\label{eq:aF12}
\end{equation}
The energy density on a slice by $1$-$2$ plane
is given as
\begin{align}
E 
&=
\int dx^2
\left[
\frac{1}{2(2g_4)^2\rho^2}
\left( 
\frac{\pa a}{\pa \rho}
\right)^2
+
f^2
\left( 
\frac{\pa \phi}{\pa \rho}
\right)^2
+
f^2
\frac{(1-a)^2}{\rho^2}
\phi^2
+
\frac{\lambda f^4}{2}
\left( 1-\phi^2 \right)^2
\right]
\nn\\
&=
\pi f^2
\int du\, u
\left[
\frac{(a')^2}{u^2}
+
2
\left( 
\phi'
\right)^2
+
2
\frac{(1-a)^2}{u^2}
\phi^2
+
\frac{\lambda}{(2g_4)^2}
\left( 1-\phi^2 \right)^2
\right]\,,
\label{eq:aEv}
\end{align}
where $u:= 2g_4 f \rho$.
From \eqref{eq:aEv},
the tension of the magnetic flux tube
is estimated as
\begin{equation}
T_{\tilde{\vec{e}}_1}
\sim
f^2 F(\lambda/(2g_4)^2)\,,
\label{eq:aT}
\end{equation}
with $F(\lambda/(2g_4)^2)$
is a function of order one.

Varying the energy functional \eqref{eq:aEv}
with respect to $\phi$ and $a$ gives
\begin{align}
\frac{d^2 a}{du^2}
-
\frac{1}{u} \frac{da}{du}
+ 
2 (1-a) \phi^2
&=0\,,
\label{eq:aeqa}\\
\frac{d^2 \phi}{du^2}
+
\frac{1}{u}
\frac{d\phi}{du}
-
\frac{(1-a)^2}{u^2} \phi
+
\frac{\lambda}{(2g_4)^2}
(1-\phi^2)
\phi
&=0\,.
\label{eq:aeqphi}
\end{align}
The set of equations \eqref{eq:aeqa} and \eqref{eq:aeqphi}
can readily be solved numerically.
One finds that the magnetic flux \eqref{eq:aF12}
rapidlly falls off beyond $\rho \sim \rho_{\mathrm{flux}}:= 1/(2 g_4 f)$.
$\rho_{\mathrm{flux}}$ can be regarded as the radius of the magnetic flux tube.

\section{$U(1)_{(k)}$ Magnetic Flux Tube}\label{sec:avk}

Below, I construct a magnetic flux solution
with $U(1)_{(j=k)}$ magnetic flux.
For this purpose,
I set the rest of the gauge fields to zero, i.e.
$A_{\mu\,(j\ne k)} = 0$.
Then, the covariant derivatives for the Higgs fields 
\eqref{eq:DH} become
\begin{align}
D_\mu H_{(k-1,k)}
&=
\pa_\mu H_{(k-1,k)} 
 + i g H_{(k-1,k)} A_{\mu (k)} \,,
\label{eq:aDHk1k}\\
D_\mu H_{(k,k+1)}
&=
\pa_\mu H_{(k,k+1)} 
- i g A_{\mu (k)} H_{(k,k+1)} \,,
\label{eq:aDHkk1}
\end{align}
I make the anzats
\begin{align}
H_{(k,k+1)}^\ast(x) = H_{(k-1,k)}(x) &= {f e^{i \theta}}\phi(g f \rho)\,,
\label{eq:aFTHs}\\
A_{\ell, (k)} (x)&= 
\epsilon_{\ell s} \hat{x}^s \frac{a(g f \rho)}{g \rho} \,.
\label{eq:aFTAk}
\end{align}
With the anzats \eqref{eq:aFTAk}, the magnetic flux is calculated as
\begin{equation}
F_{12,(k)}
=
-\frac{1}{g \rho}
\frac{\pa a}{\pa \rho} \,.
\end{equation}
The energy density on a slice by $1$-$2$ plane
is given as
\begin{align}
E 
&=
\int dx^2
\left[
\frac{1}{2g^2\rho^2}
\left( 
\frac{\pa a}{\pa \rho}
\right)^2
+
f^2
\left( 
\frac{\pa \phi}{\pa \rho}
\right)^2
+
f^2
\frac{(1-a)^2}{\rho^2}
\phi^2
+
\frac{\lambda f^4}{2}
\left( 1-\phi^2 \right)^2
\right]
\nn\\
&=
\pi f^2
\int du\, u
\left[
\frac{(a')^2}{u^2}
+
2
\left( 
\phi'
\right)^2
+
2
\frac{(1-a)^2}{u^2}
\phi^2
+
\frac{\lambda}{g^2}
\left( 1-\phi^2 \right)^2
\right]\,.
\label{eq:aEvk}
\end{align}
Here, $u:= g f \rho$.
Again, the tension of the magnetic flux tube
$T_{(k)}$ is of the order of $f^2$.

Varying the energy functional \eqref{eq:aEvk}
with respect to $\phi$ and $a$ gives
\begin{align}
\frac{d^2 a}{du^2}
-
\frac{1}{u} \frac{da}{du}
+ 
2 (1-a) \phi^2
&=0\,,
\label{eq:aeqak}\\
\frac{d^2 \phi}{du^2}
+
\frac{1}{u}
\frac{d\phi}{du}
-
\frac{(1-a)^2}{u^2} \phi
+
\frac{\lambda}{g^2}
(1-\phi^2)
\phi
&=0\,.
\label{eq:aeqphik}
\end{align}
This is the same set of equations
to \eqref{eq:aeqa} and \eqref{eq:aeqphi}
by the replacement $2g_4\rightarrow g$,
and can be readily solved numerically.
The radius of the magnetic flux 
is of the order of $\rho_{\mathrm{flux}}:=1 /(gf)$.

\bibliography{DecMono}

\providecommand{\href}[2]{#2}\begingroup\raggedright\begin{thebibliography}{10}

\bibitem{Saraswat:2016eaz}
P.~Saraswat, ``{Weak gravity conjecture and effective field theory},'' \href{http://dx.doi.org/10.1103/PhysRevD.95.025013}{{\em Phys. Rev.} {\bfseries D95} no.~2, (2017) 025013},
\href{http://arxiv.org/abs/1608.06951}{{\ttfamily arXiv:1608.06951 [hep-th]}}.

\bibitem{Hill:2000mu}
C.~T. Hill, S.~Pokorski, and J.~Wang, ``{Gauge Invariant Effective Lagrangian for Kaluza-Klein Modes},'' \href{http://dx.doi.org/10.1103/PhysRevD.64.105005}{{\em Phys. Rev.} {\bfseries D64} (2001) 105005},
\href{http://arxiv.org/abs/hep-th/0104035}{{\ttfamily arXiv:hep-th/0104035 [hep-th]}}.

\bibitem{Arkani-Hamed:2001kyx}
N.~Arkani-Hamed, A.~G. Cohen, and H.~Georgi, ``{(De)constructing dimensions},'' \href{http://dx.doi.org/10.1103/PhysRevLett.86.4757}{{\em Phys. Rev. Lett.} {\bfseries 86} (2001) 4757--4761}, \href{http://arxiv.org/abs/hep-th/0104005}{{\ttfamily arXiv:hep-th/0104005}}.

\bibitem{ArkaniHamed:2001nc}
N.~Arkani-Hamed, A.~G. Cohen, and H.~Georgi, ``{Electroweak symmetry breaking from dimensional deconstruction},'' \href{http://dx.doi.org/10.1016/S0370-2693(01)00741-9}{{\em Phys. Lett.} {\bfseries B513} (2001) 232--240},
\href{http://arxiv.org/abs/hep-ph/0105239}{{\ttfamily arXiv:hep-ph/0105239 [hep-ph]}}.

\bibitem{ArkaniHamed:2003wu}
N.~Arkani-Hamed, H.-C. Cheng, P.~Creminelli, and L.~Randall, ``{Extra natural inflation},'' \href{http://dx.doi.org/10.1103/PhysRevLett.90.221302}{{\em Phys. Rev. Lett.} {\bfseries 90} (2003) 221302},
\href{http://arxiv.org/abs/hep-th/0301218}{{\ttfamily arXiv:hep-th/0301218 [hep-th]}}.

\bibitem{Furuuchi:2020klq}
K.~Furuuchi, S.~S. Naik, and N.~J. Jobu, ``{Large Field Excursions from Dimensional (De)construction},'' \href{http://dx.doi.org/10.1088/1475-7516/2020/06/054}{{\em JCAP} {\bfseries 06} (2020) 054}, \href{http://arxiv.org/abs/2001.06518}{{\ttfamily arXiv:2001.06518 [hep-th]}}.

\bibitem{Furuuchi:2020ery}
K.~Furuuchi, N.~J. Jobu, and S.~S. Naik, ``{Extra-Natural Inflation (De)constructed},'' \href{http://arxiv.org/abs/2004.13755}{{\ttfamily arXiv:2004.13755 [hep-th]}}.

\bibitem{Poppitz:2003uz}
E.~Poppitz, ``{Deconstructing KK monopoles},'' \href{http://dx.doi.org/10.1088/1126-6708/2003/08/044}{{\em JHEP} {\bfseries 08} (2003) 044},
\href{http://arxiv.org/abs/hep-th/0306204}{{\ttfamily arXiv:hep-th/0306204 [hep-th]}}.

\bibitem{Nielsen:1973cs}
H.~B. Nielsen and P.~Olesen, ``{Vortex Line Models for Dual Strings},'' \href{http://dx.doi.org/10.1016/0550-3213(73)90350-7}{{\em Nucl. Phys.} {\bfseries B61} (1973) 45--61}.
[,302(1973)].

\bibitem{Nambu:1974zg}
Y.~Nambu, ``{Strings, Monopoles and Gauge Fields},'' \href{http://dx.doi.org/10.1103/PhysRevD.10.4262}{{\em Phys. Rev.} {\bfseries D10} (1974) 4262}.
[,310(1974); ,310(1974)].

\bibitem{Furuuchi:2017upe}
K.~Furuuchi, ``{Weak Gravity Conjecture From Low Energy Observers' Perspective},'' \href{http://dx.doi.org/10.1002/prop.201800016}{{\em Fortsch. Phys.} {\bfseries 66} no.~10, (2018) 1800016},
\href{http://arxiv.org/abs/1712.01302}{{\ttfamily arXiv:1712.01302 [hep-th]}}.

\bibitem{ArkaniHamed:2006dz}
N.~Arkani-Hamed, L.~Motl, A.~Nicolis, and C.~Vafa, ``{The String landscape, black holes and gravity as the weakest force},'' \href{http://dx.doi.org/10.1088/1126-6708/2007/06/060}{{\em JHEP} {\bfseries 06} (2007) 060},
\href{http://arxiv.org/abs/hep-th/0601001}{{\ttfamily arXiv:hep-th/0601001 [hep-th]}}.

\bibitem{Pathak:2025ukb}
M.~Pathak and K.~Furuuchi, ``{Localized Towers on the Composite Magnetic Monopole and the Weak Gravity Conjecture$=$Distance Conjecture Connection},'' \href{http://arxiv.org/abs/2506.17432}{{\ttfamily arXiv:2506.17432 [hep-th]}}.

\bibitem{Cheung:2014vva}
C.~Cheung and G.~N. Remmen, ``{Naturalness and the Weak Gravity Conjecture},'' \href{http://dx.doi.org/10.1103/PhysRevLett.113.051601}{{\em Phys. Rev. Lett.} {\bfseries 113} (2014) 051601},
\href{http://arxiv.org/abs/1402.2287}{{\ttfamily arXiv:1402.2287 [hep-ph]}}.

\bibitem{Weinberg:2012pjx}
E.~J. Weinberg, \href{http://dx.doi.org/10.1017/CBO9781139017787}{{\em {Classical solutions in quantum field theory}: {Solitons and Instantons in High Energy Physics}}}.
\newblock Cambridge Monographs on Mathematical Physics. Cambridge University Press, 9, 2012.

\bibitem{Reece:2018zvv}
M.~Reece, ``{Photon Masses in the Landscape and the Swampland},'' \href{http://dx.doi.org/10.1007/JHEP07(2019)181}{{\em JHEP} {\bfseries 07} (2019) 181},
\href{http://arxiv.org/abs/1808.09966}{{\ttfamily arXiv:1808.09966 [hep-th]}}.

\end{thebibliography}\endgroup
\bibliographystyle{utphys}
\end{document}